\documentclass[12pt]{article}

\usepackage[letterpaper, left=1in, top=1in, right=1in, bottom=1in,nohead, verbose, ignoremp]{geometry}
\usepackage{setspace}

\usepackage{times}
\usepackage{graphicx,psfrag, natbib,amsfonts,amsmath, colonequals}
\usepackage[latin1]{inputenc}
\usepackage{bm}
\usepackage{subfigure}
\usepackage{hyperref}
\usepackage{url}

\usepackage{comment}

\bibpunct[, ]{(}{)}{;}{a}{,}{,}

\newcommand{\bK}{\mathbf{W}}
\newcommand{\bl}{\mathbf{k}}
\newcommand{\knots}{\mathcal{K}}

\newcommand{\bh}{\mathbf{h}}

\newcommand{\bc}{\mathbf{c}}
\newcommand{\bb}{\mathbf{b}}
\newcommand{\bs}{\mathbf{s}}
\newcommand{\bx}{\mathbf{x}}

\newcommand{\bZ}{\mathbf{Z}}

\newcommand{\bY}{\mathbf{Y}}

\newcommand{\bI}{\mathbf{I}}

\newcommand{\bV}{\mathbf{V}}
\newcommand{\bX}{\mathbf{X}}
\newcommand{\bB}{\mathbf{B}}

\newcommand{\bfzero}{\mathbf{0}}

\newcommand{\domain}{\mathcal{D}}
\newcommand{\proposal}{\mathcal{Q}}
\newcommand{\matern}{\mathcal{M}}

\newcommand{\bfgamma}{\bm{\gamma}}
\newcommand{\bfGamma}{\bm{\Gamma}}
\newcommand{\bfmu}{\bm{\mu}}
\newcommand{\bftheta}{\bm{\theta}}
\newcommand{\bfeta}{\bm{\eta}}
\newcommand{\bfdelta}{\bm{\delta}}
\newcommand{\bfkappa}{\bm{\kappa}}
\newcommand{\bfbeta}{\bm{\beta}}
\newcommand{\bfepsilon}{\bm{\epsilon}}

\newcommand{\bfSigma}{\bm{\Sigma}}

\newcommand{\var}{var}
\newcommand{\cov}{cov}
\newcommand{\diag}{diag}

\newcommand{\GWN}{\text{GWN}}
\newcommand{\GP}{\text{GP}}

\title{Bayesian Nonstationary Spatial Modeling for Very Large Datasets}

\author{Matthias Katzfuss\thanks{Institut f\"{u}r Angewandte Mathematik, Universit\"{a}t Heidelberg, Im Neuenheimer Feld 294, 69120 Heidelberg, Germany. Email: \url{katzfuss@gmail.com}}}
\date{}

\begin{document}

\maketitle

%%%%%%%%%%%%%%%%%%%%%%%%%%%%%%%%%%%

\begin{abstract}

With the proliferation of modern high-resolution measuring instruments mounted on satellites, planes, ground-based vehicles and monitoring stations, a need has arisen for statistical methods suitable for the analysis of large spatial datasets observed on large spatial domains. Statistical analyses of such datasets provide two main challenges: First, traditional spatial-statistical techniques are often unable to handle large numbers of observations in a computationally feasible way. Second, for large and heterogeneous spatial domains, it is often not appropriate to assume that a process of interest is stationary over the entire domain.

We address the first challenge by using a model combining a low-rank component, which allows for flexible modeling of medium-to-long-range dependence via a set of spatial basis functions, with a tapered remainder component, which allows for modeling of local dependence using a compactly supported covariance function. Addressing the second challenge, we propose two extensions to this model that result in increased flexibility: First, the model is parameterized based on a nonstationary Mat\'{e}rn covariance, where the parameters vary smoothly across space. Second, in our fully Bayesian model, all components and parameters are considered random, including the number, locations, and shapes of the basis functions used in the low-rank component.

Using simulated data and a real-world dataset of high-resolution soil measurements, we show that both extensions can result in substantial improvements over the current state-of-the-art.

\vspace{.5cm}

\end{abstract}

%%%%%%%%%%%%%%%%%%%%%%%%%%%%%%%%%%%

\section{Introduction \label{introduction}}

From remote sensing of environmental variables using satellite instruments to proximal sensing of soil properties using a ground-based gamma-radiometer, a vast number of spatial measurements are now being obtained every day. Based on such very large, noisy, nongridded, and incomplete datasets, the goal is spatial prediction of a process of interest, together with rigorous quantification of prediction uncertainty. Computational feasibility for such datasets has been addressed from several angles: Approximations by Gaussian Markov random fields \citep[e.g.,][]{Lindgren2011a}, composite likelihoods \citep[e.g.,][]{Lindsay1988,Curriero1999,Bevilacqua2012,Eidsvik2012}, covariance tapering, and low-rank models. We focus here on the latter two.
% Statistical analyses of large datasets obtained over large domains (in the two examples above, the entire globe and a large farm, respectively) provide two main challenges: First, traditional spatial-statistical techniques are often unable to handle large numbers of observations (more than 10,000 or so) in a computationally feasible way. This is especially true for Bayesian models, for which posterior inference often requires computations to be carried out at each of many iterations in an MCMC sampler. The second challenge is that for large spatial domains, it is often not appropriate to assume that a process of interest is stationary over the entire domain.

Covariance tapering \citep{Furrer2006,Kaufman2008,Shaby2010} relies on compactly supported correlation functions \citep[e.g.,][]{Gneiting2002} to produce sparse covariance matrices containing only a moderate number of nonzero elements. Use of efficient sparse-matrix algorithms then may result in computational feasibility for large datasets. However, by definition, covariance tapering is most appropriate for modeling processes with weak long-range dependence.

% dimension reduction approaches
A second approach to achieving computational feasibility for large spatial datasets is through low-rank models, which include a component that can be written as a linear combination of spatial basis functions,
\begin{equation}
\label{SBF}
	\textstyle\sum_{j=1}^r b_j(\cdot)\, \eta_j = \bb(\cdot)'\bfeta,
\end{equation}
where $\bfeta | \bK \sim N_r(\bfzero,\bK)$, and the number of basis functions, $r$, is much smaller than the number of observations, $n$. Many models that include such a component have been proposed \citep[for a recent overview, see][]{Wikle2010}. The models differ in the parameterizations and priors for the covariance matrix $\bK$ and the functions in $\bb(\cdot)$. For discretized convolution models \citep[i.e., convolution models whose integrals are discretized; see, e.g.,][]{Higdon1998,Calder2007,Lemos2009}, $\bb(\cdot)$ contains the convolution kernels, and $\bK$ is often assumed to be a multiple of the identity. 
Other authors view $\bb(\cdot)$ as a vector of fixed basis functions, such as empirical orthogonal functions \citep[e.g.][]{Mardia1998,Wikle1999}, equatorial normal modes \citep[e.g.,][]{Wikle2001}, Fourier basis functions \citep[e.g.,][]{Xu2005}, W-wavelets \citep[e.g.,][]{Shi2007,Cressie2010a,Kang2009a}, or bisquare functions \citep[e.g.,][]{Cressie2008,Katzfuss2010,Katzfuss2011}. Here, we use the predictive-process approach \citep{Banerjee2008}, where both $\bb(\cdot)$ and $\bK$ are parameterized according to a ``parent process,'' for which a parametric covariance model is chosen.

Models with low-rank components \eqref{SBF} allow for fast computation via the Sherman-Morrison-Woodbury formula, as is made clear in \citet{Cressie2006} and \citet{Shi2007}. For general $\bK$, they are also flexible, in that the covariance of \eqref{SBF}, namely $\bb(\bs_1)'\bK\,\bb(\bs_2)$ for locations $\bs_1$ and $\bs_2$, is not of traditional parametric form. The fast computation and the flexibility make components of the form \eqref{SBF} very well suited to modeling medium-range to long-range spatial dependence. However, due to the dimension reduction inherent in \eqref{SBF}, a low-rank component alone is typically not able to model ``rough'' (i.e., non-smooth) short-range dependence \citep[see, e.g.,][]{Stein2008a,Finley2009}. Some efforts have been made to address this problem \citep[e.g.,][]{Wikle1999,Berliner2000,Wikle2001,Stein2008a}, including in the context of the predictive process (\citealp{Katzfuss2011d}, ch.\ 4; \citealp{Sang2011a,Sang2012}). Here, we follow the approach of \citet{Sang2012}, who divide a parent process into a predictive-process component and a remainder component. The covariance matrix of the remainder component is then made sparse by multiplication of its covariance function with a compactly supported tapering function. This approach allows for computationally feasible inference, even for large datasets.

% contribution: Nonstationary parent covariance
The contributions of this article are two extensions of the approach by \citet{Sang2012}, which allow for more flexibility and nonstationarity. First, we specify a nonstationary Mat\'{e}rn model \citep{Paciorek2006,Stein2005} for the parent covariance, in which the parameters vary smoothly across space as linear combinations of spatial basis functions.

% contribution: inference on basis functions and K
The second extension is that we allow the set of basis-function locations (henceforth referred to as ``knots'') in our low-rank component to be a random point process. This allows us to avoid choosing an arbitrary and fixed set of knots a priori. Here, $\bb(\cdot)$, $\bfeta$, and $\bK$ in \eqref{SBF} are all treated as unknown and random. This Bayesian source separation task \citep[see, e.g.,][]{Knuth2005}, where both the ``source signal'' $\bfeta$ and the ``mixing coefficients'' $\bb(\cdot)$ have to be estimated, can be achieved by putting a prior on both components. This has been done in the context of discretized-convolution models by \citet{Lemos2009}, who infer (spatially varying) parameters determining the shapes of their kernels. \citet{Lopes2008} also consider a model of the form \eqref{SBF} where both $\bb(\cdot)$ and $\bfeta$ are random, but as each basis function is itself a Gaussian process, their approach is infeasible for large spatial datasets. Recently, \citet{Guhaniyogi2011} also proposed a predictive-process model where the locations (but not the number) of the basis functions are assumed random. In this article, we implicitely make inference on the number, locations, and shapes of the basis functions. Our approach is a special case of that in \citet[][ch.\ 4]{Katzfuss2011d} and is inspired by \citet{Holmes2001}, who propose a piecewise linear spline regression model for which both the number and the locations of the splines are random.

% contribution 3
A third contribution of this article is partially philosophical in nature: We do not consider the parent process to be the truth that is to be approximated, but rather as a way of obtaining a prior for the two spatially dependent components in our model. The resulting process is more flexible than the parent process, and hence it is often preferable for modeling nonstationary real-world processes.

Posterior inference for our model is described in detail. It is fairly involved but computationally feasible, even for very large datasets. A reversible-jump Markov chain Monte Carlo algorithm \citep{Green1995} allows us to infer the number of basis functions. We take advantage of sparse-matrix operations to ensure fast computation, and we employ marginalization strategies \citep[e.g.,][]{VanDyk2008} to achieve satisfactory mixing of the Markov chain.

% organized as follows
This article is organized as follows: In Section \ref{methodology}, we introduce our nonstationary spatial model based on the model of \citet{Sang2012}. Section \ref{posteriorinference} deals with posterior inference on the unknown quantities in the model. In Section \ref{compsec}, we assess the effect of our extensions to the approach of \citet{Sang2012}, using simulated data and a real-world dataset of soil measurements. Conclusions are given in Section \ref{conclusions}.

%%%%%%%%%%%%%%%%%%%%%%
\section{Methodology \label{methodology}}

%%%%%%%
\subsection{A Standard Spatial Statistical Model \label{standardmodel}}

Let $\{ Y(\bs)\! : \: \bs \in \domain \}$, or $Y(\cdot)$, denote the process of interest on a spatial domain $\domain \subset \mathbb{R}^d$, $d \in \{1,2,3\}$. Suppose that at $n$ locations we have observations on $Y(\cdot)$, namely $Z(\bs_1),\ldots,Z(\bs_n)$, where $n$ is very large, and we assume additive measurement error:
\begin{equation}
\label{datamodeldef}
  Z(\bs_i)  \colonequals Y(\bs_i) + \epsilon(\bs_i),\quad i=1,\ldots,n,
\end{equation}
where $\epsilon(\cdot) | \sigma^2_\epsilon \sim \GWN(0,\sigma^2_\epsilon)$ is Gaussian white noise and independent of $Y(\cdot)$. For simplicity and to ensure identifiability, throughout this article we will assume that $\sigma^2_\epsilon$ is fixed and known. In practice, if $\sigma^2_\epsilon$ is not known (e.g., from instrument experiments), it can be estimated from the data by extrapolating the variogram to the origin as described in \citet{Kang2009}.

In spatial statistics, the process model is often given by,
\begin{equation}
\label{ydef}
  Y(\cdot) \colonequals \mu(\cdot) + \omega(\cdot),
\end{equation}
where $\mu(\cdot) \colonequals \bx(\cdot)'\bfbeta$ is the large-scale trend, $\bfbeta$ has an (improper) flat prior on $\mathbb{R}^p$, and $\omega(\cdot)$ is a spatially correlated component, which is typically modeled as a Gaussian process,
\begin{equation}
\label{parentprocess}
 \omega(\cdot) | \bftheta \sim \GP(0, C_P),
\end{equation}
with mean zero and covariance function
\begin{equation}
\label{parentcovariance}
 C_P(\bs_1,\bs_2) = \sigma(\bs_1) \sigma(\bs_2)\, \rho_P(\bs_1,\bs_2), \quad \bs_1, \bs_2 \in \domain,
\end{equation}
where $\sigma\!: \domain \rightarrow \mathbb{R}^+_0$ and the correlation function $\rho_P\!: (\domain \times \domain) \rightarrow [-1,1]$ are parameterized by $\bftheta$.

%%%%%%%
\subsection{A Low-Rank Component with Random Basis Functions \label{predictiveprocess}}

While the standard spatial model described in Section \ref{standardmodel} has been used extensively and successfully \citep[see, e.g.,][]{Banerjee2004}, it is computationally infeasible if $n$ is very large (more than 10,000 or so) and $C_P$ is a standard covariance function (e.g., the exponential covariance function). This is because it takes on the order of $n^3$ computations to evaluate the likelihood.

Many approximations or modeling approaches have been proposed to solve this problem (see Section \ref{introduction}). We will focus here on the predictive process \citep{Banerjee2008}. Given a so-called ``parent process'' $\omega(\cdot)$ as in \eqref{parentprocess}, the predictive process is defined as, $\nu(\cdot) \colonequals E(\omega(\cdot) | \omega(\bl_1),\ldots,\omega(\bl_r))$, where
\begin{equation}
\label{centerseq}
 \knots \colonequals \{ \bl_1,\ldots,\bl_r \}, \quad \mbox{with} \;\, \bl_j \in \domain, \, j=1,\ldots,r,
\end{equation}
is a set of knots. Conditional on $\bftheta$ and $\knots$, the predictive process can be written as a linear combination of basis functions, namely as $ \nu(\cdot)  = \bb(\cdot)'\bfeta$ with $\bfeta \sim N_r(\bfzero, \bK)$, where now
\begin{equation}
\label{BFdef}
 \bb(\bs) \colonequals \sigma(\bs) \,\big(\rho_P( \bs, \bl_1),\ldots, \rho_P( \bs, \bl_r)\big)', \quad \bs \in \domain,
\end{equation}
$\bK \colonequals  \big(( \rho_P( \bl_i, \bl_j) )_{i,j = 1,\ldots,r}\big)^{-1}$. Thus, we have $ \nu(\cdot) | \bftheta, \knots \sim \GP(0,C_\nu)$, where $C_\nu(\bs_1,\bs_2) \colonequals \bb(\bs_1)'\bK\bb(\bs_2)$, $\bs_1, \bs_2 \in \domain$.

In what follows, we do not choose a fixed set of knots $\knots$ in \eqref{centerseq}. Instead we model $\knots$ as a random point process. As discussed later at the end of Section \ref{mcmcoverview}, it is not necessary to strongly penalize large numbers of basis functions, $r$, through the prior on $\knots$. Thus, we assume a flat, noninformative, improper prior for $\knots$ with density proportional to 1.

%%%%%%%
\subsection{Adding a Tapered Remainder Component \label{fullscaleapprox}}

It was pointed out by \citet{Finley2009} that the predictive process can only account for smooth dependence. Hence, as in \citet{Sang2012}, we write:
\begin{equation}
\label{decomposition}
  \omega(\cdot) = \nu(\cdot) + (\omega(\cdot) - \nu(\cdot)) \equalscolon \nu(\cdot) + \tilde{\delta}(\cdot).
\end{equation}
Then $\tilde{\delta}(\cdot) = \omega(\cdot) - \nu(\cdot)$ is independent of $\nu(\cdot)$, and $\tilde{\delta}(\cdot) \sim \GP(0, C_{\tilde{\delta}})$, where $C_{\tilde{\delta}}(\cdot,\cdot) = C_P(\cdot,\cdot) - C_\nu(\cdot,\cdot)$ is a valid covariance function. To achieve computational feasibility for large $n$, \citet{Sang2012} proposed to replace $\tilde{\delta}(\cdot)$ in \eqref{decomposition} by $ \delta(\cdot) \sim \GP(0, C_{\delta})$, where
\begin{equation}
\label{deltacovdef}
 C_\delta ( \bs_1, \bs_2 ) = \mathcal{T} ( \| \bs_1 - \bs_2 \| / L ) \, C_{\tilde{\delta}}(\bs_1 ,\bs_2 ), \quad \bs_1, \bs_2 \in \domain
\end{equation}
is a tapered version of $C_{\tilde{\delta}}$. In \eqref{deltacovdef}, $\mathcal{T}(\cdot)$ is a compactly supported correlation function \citep[see, e.g.,][]{Gneiting2002} that is equal to zero when its argument is greater than one. Multiplication of $C_{\tilde{\delta}}$ with $\mathcal{T}$ achieves that $C_\delta( \bs_1, \bs_2 ) = 0$ if $\| \bs_1 - \bs_2 \| \geq L$, resulting in a covariance matrix that is sparse and quickly invertible (see Section \ref{computationsec} below). We will assume the tapering length $L$ to be fixed and chosen to ensure computational feasibility.

In summary, our data model is given by \eqref{datamodeldef}, and our process model is given by
\begin{equation}
\label{modelsummary}
 Y(\cdot) = \bx(\cdot)'\bfbeta + \nu(\cdot) + \delta(\cdot),
\end{equation}
where $\nu(\cdot)$ describes the medium-range to long-range spatial dependence, and $\delta(\cdot)$ accounts for local (or short-range) dependence. Both $\nu(\cdot)$ and $\delta(\cdot)$ are zero-mean Gaussian processes, whose covariance functions depend on a random set of knots, $\knots$, with a flat prior distribution, and on a parent covariance function, $C_P$, parameterized by $\bftheta$ and specified further in Section \ref{covariancesec} below.

%%%%%%%%%%%%%%%
\subsection{The Parent Covariance Function \label{covariancesec}}

Let $\matern_\upsilon$ denote the Mat\'{e}rn correlation function \citep[p.\ 50]{Stein1999}, 
\begin{equation}
\label{matern}
    \matern_\upsilon (h) = ( 2 h \sqrt{\upsilon} )^\upsilon \mathcal{K}_\upsilon( 2 h \sqrt{\upsilon} ) 2^{1-\upsilon} / \Gamma(\upsilon), \quad h > 0,
\end{equation}
and $\matern_\upsilon (0) =1$, where $ \mathcal{K}_\upsilon$ is the modified Bessel function of the second kind of order $\upsilon >0 $. Also, let 
\begin{equation}
\label{quadrformnonstationarity}
 q(\bs_1,\bs_2) = \{ 2 (\bs_1 - \bs_2)' (\bfSigma_A(\bs_1) + \bfSigma_A(\bs_2) )^{-1} (\bs_1 - \bs_2) \}^{1/2}, \quad \bs_1,\bs_2 \in \mathbb{R}^d, \, d \in \mathbb{N},
\end{equation}
be a spatially varying (SV) Mahalanobis-like distance, where $\bfSigma_A(\bs)$ is a $d \times d$ positive-definite matrix describing (local) geometric anisotropy at location $\bs$. We write, 
$\bfSigma_A(\bs) \colonequals \bm{\mathcal{R}}(\bs) \, \bfGamma(\bs) \,\bm{\mathcal{R}}(\bs)'$,
where $\bfGamma(\bs) \colonequals \diag\{\gamma_1(\bs),\ldots,\gamma_d(\bs)\}$, $\{\gamma_j\!:\: \domain \rightarrow \mathbb{R}^+, \, j=1,\ldots,d\}$ are SV scale parameters, and $\bm{\mathcal{R}}$ is a rotation matrix parameterized by SV rotation angles $\{\kappa_j\!:\: \domain \rightarrow [0, \pi/2], \, j=1,\ldots,d-1\}$. A valid nonstationary Mat\'{e}rn correlation function \citep{Paciorek2006,Stein2005} is given by,
\begin{equation}
\label{nonstatmatcor}
  \widetilde{\matern}( \bs_1, \bs_2 ) = c(\bs_1,\bs_2) \matern_{ (\upsilon(\bs_1) + \upsilon(\bs_2))/2}  (q(\bs_1,\bs_2 )), \quad \bs_1, \bs_2 \in \mathbb{R}^d, \, d \in \mathbb{N},
\end{equation}
where 
$  c(\bs_1,\bs_2) \colonequals |\bfSigma_A(\bs_1)|^{1/4} |\bfSigma_A(\bs_2)|^{1/4} | (\bfSigma_A(\bs_1) + \bfSigma_A(\bs_2))/2|^{-1/2}$.

Choosing $\rho_{P} \colonequals \widetilde{\matern}$ in \eqref{parentcovariance} results in the parent covariance
\begin{equation}
\label{parentcovdetails}
C_P(\bs_1,\bs_2) = \sigma(\bs_1) \sigma(\bs_2) \, \widetilde{\matern}(\bs_1,\bs_2),  \quad \bs_1, \bs_2 \in \domain \subset \mathbb{R}^d, \; d \in \{1,2,3\}.
\end{equation}
This nonstationary Mat\'{e}rn class is very flexible, in that it allows for SV standard deviation $\sigma(\cdot)$, SV smoothness parameter $\upsilon(\cdot)$, and SV geometric anisotropy through SV scale parameters $\{\gamma_j(\cdot)\!:\: j=1,\ldots,d\}$ and SV rotation angles $\{\kappa_j(\cdot)\!:\: j=1,\ldots,d-1\}$.

To ensure computational feasibility, we let the parameters vary spatially according linear combinations of spatial basis functions. We assume that all SV parameters are determined by the (random) parameter vector, $\bftheta \colonequals (\tilde{\sigma}, \bfeta_{\sigma}',\tilde{\upsilon}, \bfeta_{\upsilon}',\tilde{\bfgamma}', \bfeta_{\bfgamma}', \tilde{\bfkappa}', \bfeta_{\bfkappa}')'$, through models of the form,
\begin{equation}
\label{generalcovparam}
  \theta(\bs) = g_{\theta}(\tilde{\theta} + \bb_\theta(\bs)'\bfeta_{\theta}), \quad \bs \in \domain,
\end{equation}
where $\theta(\cdot)$ is a generic notation for one of the SV parameters, $\tilde{\theta} \sim N(\mu_{\theta}, \sigma^2_{\theta})$, $\bfeta_{\theta} \sim N_{r_\theta} ( \bfzero, \tau^2_{\theta} \bI_{r_\theta})$, and $\bb_\theta(\cdot)$ is an $r_\theta$-dimensional vector of \emph{fixed} basis functions (same for all parameters), each normalized to $[0,1]$. The functions $g_{\theta}(\cdot)$ are transformations from $\mathbb{R}$ to the range of $\theta(\cdot)$.

\begin{table}[ht]
\begin{center}
\caption{\label{covparamtransformations}Details for the SV covariance parameters of the form \eqref{generalcovparam}}
\begin{tabular}{| l | l | l | l | l | l |}
\hline
Parameter	& Symbol $\theta(\cdot)$	& Range of $\theta(\cdot)$ & Transformation $g_{\theta}(\cdot)$ & $\mu_{\theta}$ & $\sigma^2_{\theta}$\\
\hline
Standard deviation &	$\sigma(\cdot)$ &	$\mathbb{R}^+$	& $\exp(\cdot)$ & $(^*)$ & $\sigma^2_\sigma = 0.25$\\
Smoothness	& $\upsilon(\cdot)$ &	$ [0, 2]$	& $2\Phi(\cdot)$ & $\mu_\upsilon = 0$ & $\sigma^2_\upsilon =1$\\
Scale	& $\gamma_j(\cdot)$	& $\mathbb{R}^+$	& $\exp(\cdot)$ & $(^*)$ & $\sigma^2_\gamma = 0.25$\\
Rotation angle	& $\kappa_j(\cdot)$	& $ [0, \pi/2]$	& $(\pi/2) \Phi(\cdot)$ & $\mu_\kappa = 0$ & $\sigma^2_\kappa =1$\\
\hline
\end{tabular}
\end{center}
\vspace{-.4cm}
$\Phi(\cdot)$: Cumulative distribution function of the standard normal distribution. \\
$(^*)$: The prior means $\mu_\sigma$ and $\mu_\gamma$ depend on the application; see Section \ref{compsec} for specific choices.
\end{table}

Specific choices for $g_{\theta}(\cdot)$, $\mu_{\theta}$, and $\sigma^2_{\theta}$ are given in Table \ref{covparamtransformations}. For example, we have $\sigma(\bs) = \exp(\tilde{\sigma} + \bb_\theta(\bs)'\bfeta_\sigma)$, $\tilde{\sigma} \sim N(\mu_\sigma,\sigma^2_\sigma\!=\! 0.25)$, and $\bfeta_\sigma \sim N_{r_\theta} ( \bfzero, \tau^2_{\theta} \bI_{r_\theta})$. Note that we restrict the smoothness parameter $\upsilon(\cdot)$ to the interval $[0,2]$, as ``the data can rarely inform about smoothness of higher orders'' \citep{Banerjee2008}. The parameter $\tau^2_{\theta}$ determines how much $\theta(\cdot)$ is allowed to vary \emph{a priori} over $\domain$; we set $\tau^2_{\theta} = (0.25)^2$ for all SV parameters \citep[see][]{Katzfuss2011}, inducing shrinkage towards stationarity for the covariance function $C_P$.

For $\bb_\theta(\cdot)$ in \eqref{generalcovparam}, any choice of basis functions is possible. Assuming that the covariance parameters vary smoothly over space, we choose a relatively small number of power exponential correlation functions, $\bb_\theta(\bs) = (\exp\{-((\bs- \bc_1)/\lambda)^2\},\ldots,\exp\{-((\bs - \bc_{r_\theta})/\lambda)^2\})'$, with (relatively large) fixed scale parameter $\lambda$, and fixed centers $\bc_1,\ldots,\bc_{r_\theta}$. Specific choices depend on the domain $\domain$ and are given in Section \ref{compsec}.

Our choice for $\mathcal{T}$ in \eqref{deltacovdef} in this article is Kanter's function \citep{Kanter1997}:
\begin{equation}
\label{kantersfct}
 \mathcal{T} (x) \colonequals \textstyle(1-x) \frac{\sin(2 \pi x )}{2 \pi x} + \frac{1 - \cos( 2 \pi x)}{2 \pi^2 x}, \quad x \in (0, 1),
\end{equation}
$\mathcal{T} (x) \colonequals 0$ for $x \geq 1$, and we set $\mathcal{T} (0)\colonequals 1$. The function $\mathcal{T}(\|\bh\|)$ is positive-definite for $\bh \in \mathbb{R}^3$, it is twice differentiable at the origin, and it minimizes the curvature at 0 within the class of all compactly supported and valid (in $\mathbb{R}^3$) correlation functions \citep{Gneiting2002}.

In summary, for fixed $\bfbeta$, $\knots$, and $\bftheta$, the covariance function of the true process $Y(\cdot)$ in \eqref{modelsummary} is
\begin{equation}
\label{Ycovariance}
 C_Y(\bs_1,\bs_2) = C_\nu(\bs_1,\bs_2) + \mathcal{T} ( \| \bs_1 - \bs_2 \| / L ) \{ C_P(\bs_1,\bs_2) - C_\nu(\bs_1,\bs_2) \},  \quad \bs_1, \bs_2 \in \domain,
\end{equation}
where $C_\nu$ and $\mathcal{T}$ are given by \eqref{BFdef} and \eqref{kantersfct}, respectively.
It follows immediately from Proposition 1 in \citet{Sang2012} that this covariance function is positive definite. It is a close approximation to $C_P(\cdot,\cdot)$ for a large, dense set of knots, $\knots$ (or for large $L$). Here, because $\knots$ is random, \eqref{Ycovariance} is more flexible than the parent covariance and hence preferable in many nonstationary real-world situations. Note that, because $\sigma(\cdot)$ is infinitely differentiable, $T(\cdot)$ is twice differentiable at the origin, and $\widetilde{\matern}(\bs,\bs+\bh)$ is also at most twice differentiable for $\upsilon(\bs)<2$ \citep[see also][]{Paciorek2006}, the smoothness of $Y(\cdot)$ at location $\bs \in \domain$ is solely determined by $\upsilon(\bs)$ (for fixed $\bfbeta$, $\knots$, and $\bftheta$).

%%%%%%%%%%%%%%%%%%%%%%%%%%%%%%%%%%%%%%%
\section{Posterior Inference \label{posteriorinference}}

\subsection{Summary of the Model in Vector Notation \label{sec:modelsummary}}

Integrating out $\bfeta$ and $\delta(\cdot)$, the data, $\bZ \colonequals (Z(\bs_1),\ldots,Z(\bs_n))'$, are distributed as, $  \bZ | \Omega \sim N_n( \bX \bfbeta, \bfSigma_Z)$,
where $\Omega \colonequals \{\bfbeta, \bftheta, \knots\}$, and the $i$-th row of $\bX$ is given by $\bx(\bs_i)'$. The data covariance matrix is,
\begin{equation}
\label{datacov}
\bfSigma_Z \colonequals \var( \bZ | \Omega ) = \bB \bK \bB' + \bV,
\end{equation}
where the $i$-th row of the $n \times r$ matrix $\bB$ is given by $\bb(\bs_i)'$ (see \eqref{BFdef}), $\bK$ is defined below \eqref{BFdef}, $\bV \colonequals \bV_\delta + \bV_\epsilon$, $\bV_\epsilon \colonequals \sigma^2_\epsilon \bI_n$, and $\bV_\delta \colonequals ( C_\delta(\bs_i, \bs_j))_{i,j = 1,\ldots,n} $ is the sparse $n \times n$ covariance matrix of the vector $\bfdelta \colonequals (\delta(\bs_1), \ldots, \delta(\bs_n))'$ (see \eqref{deltacovdef}).

In what is to follow, $[A]$ will denote the distribution or density of a generic random variable $A$, and $[A\, | \,\ldots\,]$ will denote the full conditional distribution of $A$ (i.e., the distribution of $A$ given the data and all parameters other than $A$ in $\Omega$). Further, let $N_k(\mathbf{a}|\bfmu,\bfSigma)$ denote the probability density function of a $k$-variate normal distribution with mean $\bfmu$ and covariance matrix $\bfSigma$, evaluated at $\mathbf{a}$. 

The densities of the full conditional distributions of the elements of $\Omega$ are all proportional to,
\[
 [\bZ,\Omega] = [\bZ | \Omega] [\Omega] = N_n(\bZ | \bX \bfbeta, \bfSigma_Z) [\bfbeta] [\bftheta] [\knots],
\]
where $[\knots] \propto 1$, $[\bfbeta] \propto 1$, and $[\bftheta]$ is described below \eqref{generalcovparam}.

\subsection{The Reversible-Jump MCMC Algorithm \label{mcmcoverview}}

For posterior inference, we will employ a reversible jump Markov chain Monte Carlo (MCMC) algorithm \citep{Green1995} based on a Gibbs sampler \citep{Geman1984} with some adaptive Metropolis-Hastings steps \citep{Metropolis1953, Hastings1970,Haario2001}. We will emphasize dependence of $\bfSigma_Z$ on a set of parameters by placing the parameters in parentheses.

The MCMC sampler consists of the following steps:
\begin{enumerate}
 \item Sample $\bfbeta$ from, 
$  [\bfbeta |\,\ldots\,] = N_p\big( \bfbeta | (\bX' \bfSigma_Z^{-1} \bX)^{-1} \bX' \bfSigma_Z^{-1} \bZ, (\bX' \bfSigma_Z^{-1} \bX)^{-1}\big)$.
\item Sample $\bftheta$ using a Metropolis-Hastings step from, 
$	[\bftheta | \,\ldots\, ] \propto  [\bftheta ] \, N_n\big(\bZ | \bX \bfbeta, \bfSigma_Z(\bftheta)\big)$.
\item Sample a new set of knots from $ [ \knots | \,\ldots\, ] $, as follows. At each MCMC iteration, we propose one of three modifications to the current set of knots, each with probability $1/3$:
\begin{enumerate}
 \item \label{addBF} Add a knot: Draw a new knot, $\bl_{r+1}$, from a uniform distribution on $\domain$, and let $\knots^{*} \colonequals \knots \cup \{ \bl_{r+1}\}$ be the proposed set of knots, which now has size $r^{*} = r+1$.
  \item \label{delBF} Delete a knot: Select one knot uniformly at random from $\knots$; that is, draw $J \sim U(1,2,\ldots, r)$. Then set $\knots^{*} \colonequals \knots \backslash \{ \bl_J \}$ and $r^{*} = r-1$. 
 \item Moving a knot (a combination of (a) and (b)): First select a knot uniformly at random to be deleted, and then select a location uniformly on $\domain$ at which to add a new one (i.e., where to move the old knot). This results in $\knots^{*} \colonequals \{ \bl_{r+1}\} \cup \knots \backslash \{ \bl_J \}$ and $r^{*} = r$.
\end{enumerate}
The reversible-jump acceptance probability \citep[][]{Green1995} for the proposed $\knots^{*}$ can be shown to be equal to $ \min\{1,\alpha\}$, where
\begin{equation}
\label{centersacceptance}
  \alpha \colonequals \frac{N_n(\bZ | \bX \bfbeta, \bfSigma_Z(\knots^{*})) }{N_n(\bZ | \bX \bfbeta, \bfSigma_Z(\knots) )} \frac{\proposal( \knots^{*}, \knots)}{\proposal( \knots, \knots^{*})},
\end{equation}
and the proposal ratio is given by,
\begin{equation}
\label{proposalratio}
 \frac{\proposal_{\knots}( \knots^{*}, \knots)}{\proposal_{\knots}( \knots, \knots^{*} )} \colonequals \left\{ \begin{array}{l l} 
1/(r+1), & r^{*} = r+1\\
r, & r^{*} = r-1\\
1, & r^{*} = r . \end{array} \right.
\end{equation}
Note that for $r=0$, deleting or moving a basis function is impossible, and so in this case we always propose to add a basis function. As a result, the proposal ratio in \eqref{proposalratio} is given by $1/3$ when $r=0$.
\end{enumerate}

There might be a concern that, for very large datasets, the data always favor a very large number of basis functions, unless there is strong penalization for large $r$ through the prior distribution on $\knots$. However, note that the acceptance probability \eqref{centersacceptance} for a proposed set of knots, $\knots^*$, is the product of the Bayes factor (of $\knots^*$ versus $\knots$) and a term depending only on the proposal distribution chosen for $\knots^*$ \citep[cf.][App.\ I]{Holmes2000}. This is reassuring, as ``the Bayes factor functions as a fully automatic Occam's razor'' \citep[][p.\ 790]{Kass1995}, and so there is strong intuition that our flat prior, $[\knots] \propto 1$, is sufficient and that no explicit penalty for large $r$ is necessary.

\subsection{Spatial Prediction \label{spatialprediction}}

In spatial statistics, the main interest is typically in making inference on the true process $Y(\cdot)$ at a set of prediction locations, $\{\bs_1^P,\ldots, \bs_{n_P}^P\}$, which might or might not include the set of observed locations. We write, 
 $ \bY^P = \bX^P \bfbeta + \bB^P \bfeta + \bfdelta^P$,
and so we need samples from,
\begin{equation}
\label{spatialpredeq}
 [  \Omega, \bfeta, \bfdelta^P | \bZ ] =  [ \Omega | \bZ ] \, [\bfeta | \Omega, \bZ] \, [ \bfdelta^P | \bfeta, \Omega, \bZ],
\end{equation}
where samples of the first term on the right-hand side were obtained in Section \ref{mcmcoverview}. Because it can be very computationally expensive, we only obtain samples of $\bfeta$ and $\bfdelta^P$ for thinned MCMC iterations after convergence of the MCMC for $\Omega$ \citep[see][for why this is valid]{VanDyk2008}. We have
\[
 \bfeta | \Omega, \bZ \sim N_r \left( (\bB' \bV^{-1} \bB + \bK^{-1} )^{-1} \bB' \bV^{-1}(\bZ - \bX \bfbeta),(\bB' \bV^{-1} \bB + \bK^{-1} )^{-1} \right)
\]
and
\begin{equation}
\label{deltafullconditional}
   \bfdelta^P | \bfeta, \Omega, \bZ \sim N_{n_P} \left( \bV_\delta^{P,O} \bV^{-1} (\bZ - \bX \bfbeta - \bB \bfeta), \bV_{\delta}^P - \bV_\delta^{P,O} \bV^{-1} \bV_\delta^{P,O}{}'\right),
\end{equation}
where $\bV_\delta^P \colonequals \var(\bfdelta^P)$, $\bV_\delta^{P,O} \colonequals \cov ( \bfdelta^P, \bfdelta )$, and $\bfdelta \colonequals (\delta(\bs_1),\ldots,\delta(\bs_n))'$. After appropriate reordering, we write $\bfdelta^P = [\bfdelta', \bfdelta^U{}']'$, where $\bfdelta^U$ denotes $\delta(\cdot)$ evaluated at all unobserved prediction locations. 
To avoid having to obtain $\bV_\delta^{P,O} \bV^{-1} \bV_\delta^{P,O}{}'$ explicitly, we obtain a sample from \eqref{deltafullconditional} by calculating the quantity,
$
  \check{\bfdelta}^P + \bV_\delta^{P,O} \bV^{-1} (\bZ - \bX \bfbeta - \bB \bfeta - \check{\bfdelta} - \check{\bfepsilon}),
$
where $\check{\bfdelta}^P \colonequals (\check{\bfdelta}', \check{\bfdelta}^U{}')' \sim N_{n_P}(\bfzero,\bV_\delta^P)$ and $\check{\bfepsilon} \sim N_n(\bfzero, \bV_\epsilon)$ \citep[cf. conditional simulation,][Sect.\ 3.6.2]{Cressie1993}.

\subsection{Computational Issues \label{computationsec}}

Note that $\bfSigma_Z$ in \eqref{datacov} is a dense $n \times n$ matrix of full rank $n$, and so naive calculation of its inverse, which appears in the MCMC updates, is computationally infeasible for large $n$. However, we can employ the Sherman-Morrison-Woodbury formula \citep{Sherman1950,Woodbury1950,Henderson1981} to obtain, $
 \bfSigma_Z^{-1} = \bV^{-1} - \bV^{-1} \bB (\bK^{-1} + \bB' \bV^{-1} \bB)^{-1} \bB' \bV^{-1},
$
and a similar formula gives, $ | \bfSigma_Z | = | \bV | | \bI_r + \bK \bB' \bV^{-1} \bB |$ \citep[e.g.,][]{Cressie2008}.
Since the tapering range, $L$ in \eqref{deltacovdef}, is fixed, the position of the nonzero elements of $\bV$ is the same for all MCMC iterations. Hence, we order the locations to allow for efficient Cholesky decomposition of $\bV$ (e.g., using the minimum-degree ordering) only once, at the beginning of the MCMC algorithm. 

In general, the number of computations required for operations involving a sparse matrix depends on the number and locations of the nonzero elements \citep{Gilbert1992}. Some numerical results in \citet{Furrer2006} indicate that the time required to compute the Cholesky decomposition of a tapered $n \times n$ covariance matrix increases roughly linearly with $n$ (for fixed domain, fixed tapering length, and a regular sampling grid), which in turn indicates that the computational complexity of our algorithm is approximately of order $n$. Questions about theoretical computational complexity aside, in our experience the majority of computation time at each of the MCMC iterations was actually spent on evaluating the modified Bessel function in \eqref{matern} for each of the nonzero elements of the matrix $\bV_\delta$ (and of $\bV_{\delta}^P$ and $\bV_\delta^{P,O}$ for iterations in which spatial predictions are obtained). We have considerable control over the speed of the MCMC algorithm through selection of the tapering range, $L$ in \eqref{deltacovdef}. For extremely massive datasets, we can set $L$ to a very small value to achieve computational feasibility.

%%%%%%%%%%%%%%%%%%%%%%%%%%%%%
\section{Numerical Model Comparisons \label{compsec}}

In this section, we will compare our model to the model of \citet{Sang2012}, which represents the current state-of-the-art in terms of geostatistical approaches to the analysis of large spatial datasets. \citet{Sang2012} showed that their model can result in better predictions and model fit than the predictive-process approach of \citet{Banerjee2008}. Our approach can be viewed as an extension of the \citet{Sang2012} model in terms of two components: random knots and the use of the nonstationary Mat\'{e}rn covariance of Section \ref{covariancesec} as the parent covariance function. Therefore, our comparisons will examine the effects of two factors: random versus (a varying number of) fixed knots, and a nonstationary Mat\'{e}rn parent covariance (NPC) versus a stationary one (SPC). (SPC is a special case of our model obtained by setting $\tau^2_\theta =0$ in \eqref{generalcovparam}.)

%%%%%%%%%%%%%%%
\subsection{Simulation Studies in One Spatial Dimension \label{simstudy}}

For the following three simulation studies, the true process is assumed to exist on a one-dimensional domain, $\domain = [1,512]$, with potential measurement locations at $\{1,2,\ldots,512\}$.

\begin{figure}
\centering\includegraphics[width=1\textwidth]{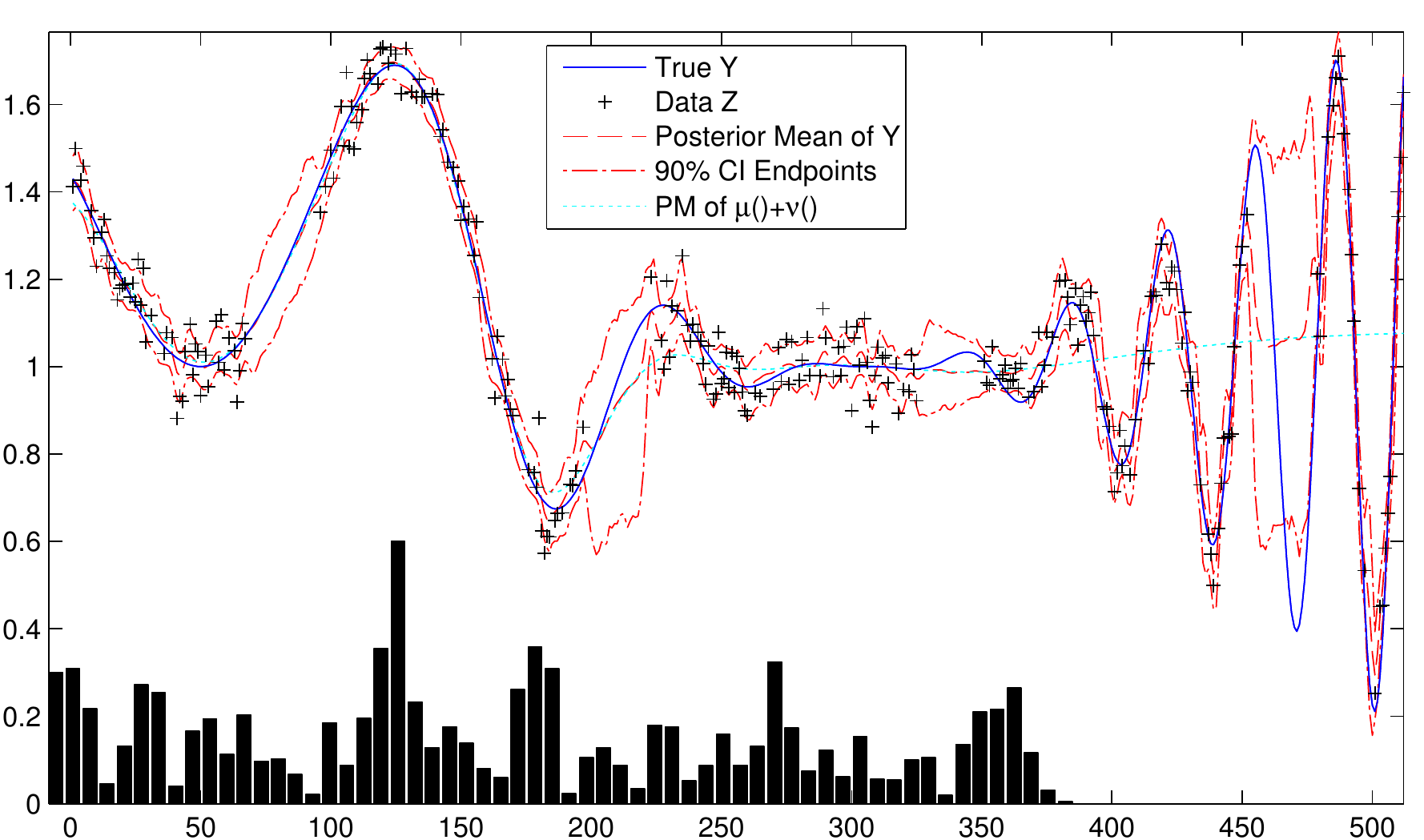}
\caption{\label{sim_example}For Simulation Study 1, the true process $Y(\cdot)$ and one sample of the data $\bZ$, together with the posterior mean and a point-wise posterior $90\%$ credible interval (CI) of $Y(\cdot)$, the posterior mean (PM) of $\mu(\cdot)+\nu(\cdot)$ (i.e., without $\delta(\cdot)$), and the density of the knots (histogram at the bottom) using our model} 
\end{figure}

In Simulation Study 1, we assumed that the true process $Y(\cdot)$ is a deterministic function:
\begin{equation}
	\label{sim1truth}
	\textstyle Y(s) = 1 + \sin\left(2\pi \left(\frac{s-306}{512}\right)^2 \right) \sin\left(20\pi \left(\frac{s-50}{512}\right)^2\right), \quad s \in \domain.
\end{equation}
This true process is shown in Figure \ref{sim_example}.

Based on this true process, we created 100 datasets of observations by adding independent normal measurement error with variance $\sigma^2_\epsilon = \hat{\sigma}^2_Y \cdot 5\% =  0.004$, where $\hat{\sigma}^2_Y = 0.08 $ is the empirical variance of $\{Y(1),\ldots,Y(512)\}$. To examine the medium-to-long-range prediction performance of the models, we created four test intervals, in which no data was observed (collectively referred to as missing by design, or MBD). These test intervals each have length 25 and begin at locations 70, 198, 326, 454, respectively. In addition, one third of the remaining locations (henceforth referred to as missing at random, or MAR) were selected at random at each iteration of the simulation study as unobserved test locations (to test short-range prediction performance near observed locations). The remaining 275 observed locations will be denoted OBS. 

To ensure comparability of the results, we assumed the measurement-error variance to be known for all models. For each of the 100 simulated datasets, each of the models was run for 10,000 MCMC iterations (thinned by a factor of 10), the first 5,000 of which were taken as burn-in (based on examination of trace plots in a pilot study). The tapering length in \eqref{deltacovdef} was chosen as $L = 6.5$, resulting in about 2,400 nonzero elements (less than 9 per row) for $\bV$ in \eqref{datacov}. The prior distributions of the parameters of the parent covariance were as described in Section \ref{covariancesec}, with $\mu_{\sigma}=\log ( \hat{\sigma}_Y )$ and $\mu_{\gamma}=\log ( 3000 )$. The spatial trend, $\mu(\cdot)$ in \eqref{ydef}, consisted only of an intercept (i.e., $\bx(\cdot) \equiv 1$).

For the random knots, the proposal distribution for new knots (as described in step 3 of Section \ref{mcmcoverview}) was a uniform distribution on $[-9, 522]$. As a pilot study using our model showed that the posterior mean of $r $ was around $11$, we used two sets of fixed knots: The first consisted of eight evenly spaced knots between locations -10 and 522, and the second set consisted of 14 evenly spaced knots between -4 and 516.
For the models with nonstationary parent covariance, we took $\bb_\theta(\cdot)$ in \eqref{generalcovparam} to be made up of four power exponential functions with scale parameter $\lambda = 74$, centered at locations 64, 192, 320, 448, respectively.  One set of observations, $\bZ$, is shown in Figure \ref{sim_example}, together with a summary of the corresponding results using our model. Very few knots are selected between locations 380 and 500, because the process fluctuates so quickly in that area that it can basically be picked up entirely by the tapered remainder component, $\delta(\cdot)$.

To measure prediction accuracy of the models under consideration, we used the mean squared prediction error, the squared difference between the true process $Y(\cdot)$ and the posterior mean for each of the models. To quantify the accuracy of the uncertainty estimation, we also calculated the interval score, which combines the width of a credible interval (here, $95\%$ posterior credible intervals) with a penalty for not containing the true value \citep[see][Sect.\ 6.2]{Gneiting2007}. The goal is for a small interval score. Both mean squared prediction error and interval score were averaged over the 100 simulated datasets and all 512 locations (ALL), and also averaged within each of the groups of locations described earlier (OBS, MAR, MBD).

\begin{table}[ht]
\begin{center}
\caption{\label{sim_results_sinefun}Results of Simulation Study 1}
\begin{tabular}{| l | r r | r r | r r | r r | r r | r r | }
\hline
& \multicolumn{2}{|c|}{Random knots} & \multicolumn{2}{|c|}{8 Fixed knots} & \multicolumn{2}{|c|}{14 Fixed knots} & \multicolumn{2}{|c|}{Full model} \\
Parent covariance	&	\multicolumn{1}{c}{NPC}	&	\multicolumn{1}{c|}{SPC}	&	\multicolumn{1}{c}{NPC}	&	\multicolumn{1}{c|}{SPC}	&	\multicolumn{1}{c}{NPC}	&	\multicolumn{1}{c|}{SPC}	&	\multicolumn{1}{c}{NPC}	&	\multicolumn{1}{c|}{SPC}	\\
\hline																	
Time (min)	&	3.64	&	3.79	&	2.01	&	2.01	&	2.76	&	2.73	&	101.17	&	100.05 \\
\hline																
MSPE (ALL) $\times 100$	&	1.00	&	1.02	&	1.19	&	1.25	&	1.38	&	1.56	&	1.48	&	1.57 \\
MSPE (OBS) $\times 100$	&	0.11	&	0.11	&	0.18	&	0.21	&	0.13	&	0.19	&	0.14	&	0.18 \\
MSPE (MAR) $\times 100$	&	0.24	&	0.25	&	0.51	&	0.57	&	0.36	&	0.48	&	0.23	&	0.28 \\
MSPE (MBD) $\times 100$	&	4.48	&	4.58	&	4.87	&	5.05	&	6.24	&	6.80	&	6.88	&	7.17 \\
\hline																
IS (ALL) $\times 100$	&	26.71	&	28.45	&	33.16	&	47.28	&	45.65	&	72.84	&	62.94	&	80.60 \\
IS (OBS) $\times 100$	&	15.54	&	15.77	&	20.11	&	21.77	&	17.11	&	20.99	&	18.30	&	22.42 \\
IS (MAR) $\times 100$	&	21.64	&	21.93	&	33.01	&	38.56	&	27.16	&	36.52	&	23.92	&	29.77 \\
IS (MBD) $\times 100$	&	64.40	&	72.26	&	69.23	&	129.36	&	149.43	&	265.21	&	239.18	&	310.26 \\
\hline																
Posterior mean of $r$	&	10.59	&	11.24	&	(8)	&	(8)	&	(14)	&	(14)	&	(275)	&	(275	)\\
\hline
\end{tabular}
\vspace{1pt}

NPC = nonstationary parent covariance; SPC = stationary parent covariance; MSPE = mean square prediction error; IS = interval score; Time = average time for each MCMC (averaged over the 100 simulated datasets); ALL = all 512 locations; OBS = the 275 observed locations; MAR = the 137 missing-at-random locations; MBD = the 100 missing-by-design locations
\end{center}
\end{table}

The results for Simulation Study 1 are shown in Table \ref{sim_results_sinefun}. Two trends are evident in terms of both scores: Using a NPC produced better predictions than using a SPC, and random knots resulted in better predictions when compared to fixed knots. The more fixed knots were used (we even included the full model, for which $r = n$), the closer the resulting models were to their respective parent processes (which are clearly the wrong models for $Y(\cdot)$ in \eqref{sim1truth}), and the worse the scores were for the test regions (MBD).

Throughout this article, we assumed that most real-world processes do not have covariances of simple parametric form. To examine the performance of our model in the unlikely event of encountering a process that does exhibit simple parametric covariance, we conducted two more simulation studies. In Simulation Studies 2 and 3, we sampled a new true process $Y(\cdot)$ 100 times each as a constant spatial ``trend'' equal to 1 plus a mean-zero Gaussian process component with the Mat\'{e}rn covariance function of \eqref{parentcovdetails}. For Simulation Study 2, we chose
\begin{equation}
\label{simstudyparams}
\begin{split}
\sigma(s) & = 3 \exp\big(\sin( (1- |s /256 -1 |) \,2 \pi )/2\big)  \\
\gamma(s) &=600 \exp\big( - 2\sin (s \,2 \pi /256)\big) (s/256)\\
\upsilon(s) & = 3 \Phi\big( - \sin (s \, 2 \pi /256)\big),
\end{split}
\end{equation}
and for Simulation Study 3, we used a stationary Mat\'{e}rn covariance with $\sigma(s) \equiv 3$, $\gamma(s) \equiv 600$, and $\upsilon(s) \equiv 1$. At each of the 100 iterations, we then simulated data, $\bZ$, by adding a spatially independent measurement-error term with variance $\sigma^2_\epsilon = 3^2 \cdot 5\% = 0.45$ at each observed location. The remaining setup was exactly the same as in Simulation Study 1, except that we chose $\mu_{\sigma}=\log (3 )$ and $\mu_{\gamma}=\log ( 600 )$.

\begin{table}[ht]
\begin{center}
\caption{\label{sim_results_nonstatmat}Results of Simulation Study 2}
\begin{tabular}{| l | r r | r r | r r | }
\hline
& \multicolumn{2}{|c|}{Random knots} & \multicolumn{2}{|c|}{8 Fixed knots} & \multicolumn{2}{|c|}{14 Fixed knots} \\
Parent covariance	&	\multicolumn{1}{c}{NPC}	&	\multicolumn{1}{c|}{SPC}	&	\multicolumn{1}{c}{NPC}	&	\multicolumn{1}{c|}{SPC}	&	\multicolumn{1}{c}{NPC}	&	\multicolumn{1}{c|}{SPC}	\\
\hline															
Time (sec)	&	184.12	&	192.42	&	111.22	&	110.38	&	152.10	&	150.19	\\
\hline													
MSPE (ALL)	&	1.80	&	1.89	&	2.02	&	2.14	&	1.95	&	2.07	\\
MSPE (OBS)	&	0.23	&	0.28	&	0.26	&	0.35	&	0.24	&	0.34	\\
MSPE (MAR)	&	1.66	&	1.84	&	1.69	&	1.73	&	1.61	&	1.70	\\
MSPE (MBD)	&	6.30	&	6.38	&	7.28	&	7.61	&	7.12	&	7.31	\\
\hline													
IS (ALL)	&	4.76	&	5.83	&	4.97	&	7.28	&	4.82	&	7.46	\\
IS (OBS)	&	2.28	&	2.69	&	2.40	&	2.93	&	2.30	&	2.89	\\
IS (MAR)	&	5.65	&	7.74	&	5.45	&	8.44	&	5.08	&	8.43	\\
IS (MBD)	&	10.37	&	11.87	&	11.39	&	17.68	&	11.38	&	18.67	\\
\hline													
Posterior mean of $r$	&	8.82	&	9.78	&	(8)	&	(8)	&	(14)	&	(14)	\\
\hline
\end{tabular}
\vspace{1pt}

NPC = nonstationary parent covariance; SPC = stationary parent covariance; MSPE = mean square prediction error; IS = interval score; Time = average time for each MCMC (averaged over the 100 simulated datasets); ALL = all 512 locations; OBS = the 275 observed locations; MAR = the 137 missing-at-random locations; MBD = the 100 missing-by-design locations
\end{center}
\end{table}

In Simulation Study 2 (see Table \ref{sim_results_nonstatmat}), the NPC models worked better than the corresponding SPC models (as expected, because the true $Y(\cdot)$ was nonstationary). Overall, random knots resulted in better predictions than fixed knots, especially for the (misspecified) SPC models.

\begin{table}[ht]
\begin{center}
\caption{\label{sim_results_statmat}Results of Simulation Study 3}
\begin{tabular}{| l | r r | r r | r r | }
\hline
& \multicolumn{2}{|c|}{Random knots} & \multicolumn{2}{|c|}{8 Fixed knots} & \multicolumn{2}{|c|}{14 Fixed knots}  \\
Parent covariance	&	\multicolumn{1}{c}{NPC}	&	\multicolumn{1}{c|}{SPC}	&	\multicolumn{1}{c}{NPC}	&	\multicolumn{1}{c|}{SPC}	&	\multicolumn{1}{c}{NPC}	&	\multicolumn{1}{c|}{SPC}	\\
\hline															
Time (sec)	&	193.68	&	196.75	&	108.87	&	109.51	&	149.12	&	148.85	\\
\hline													
MSPE (ALL)	&	1.11	&	1.12	&	1.56	&	1.58	&	1.24	&	1.25	\\
MSPE (OBS)	&	0.20	&	0.20	&	0.24	&	0.24	&	0.22	&	0.22	\\
MSPE (MAR)	&	0.40	&	0.40	&	0.55	&	0.57	&	0.47	&	0.48	\\
MSPE (MBD)	&	4.58	&	4.62	&	6.57	&	6.67	&	5.08	&	5.10	\\
\hline													
IS (ALL)	&	4.12	&	4.20	&	5.07	&	5.16	&	4.68	&	4.71	\\
IS (OBS)	&	2.13	&	2.14	&	2.30	&	2.31	&	2.25	&	2.26	\\
IS (MAR)	&	3.12	&	3.14	&	3.62	&	3.67	&	3.38	&	3.38	\\
IS (MBD)	&	10.93	&	11.33	&	14.69	&	15.04	&	13.13	&	13.27	\\
\hline													
Posterior mean of $r$	&	10.40	&	10.69	&	(8)	&	(8)	&	(14)	&	(14)	\\
\hline
\end{tabular}
\vspace{1pt}

NPC = nonstationary parent covariance; SPC = stationary parent covariance; MSPE = mean square prediction error; IS = interval score; Time = average time for each MCMC (averaged over the 100 simulated datasets); ALL = all 512 locations; OBS = the 275 observed locations; MAR = the 137 missing-at-random locations; MBD = the 100 missing-by-design locations
\end{center}
\end{table}

For Simulation Study 3 (see Table \ref{sim_results_statmat}), NPC still worked slightly better than SPC, suggesting that there is no penalty in terms of predictive distributions for using the (more flexible) NPC model when the true process is stationary. The models with random knots again had the best results.

For all three simulation studies, the models with random knots resulted in longer computation times than the models with eight or 14 fixed knots.

%%%%%%%%%%%
\subsection{Analysis of Soil Readings from a Gamma-Radiometer \label{application}}

We also compared the models of Section \ref{simstudy} using a large real-world spatial dataset. \citet{ViscarraRossel2007} collected high-resolution soil information on Nowley farm in New South Wales, Australia. It is important to develop automated soil sensing for monitoring and precision agriculture, because conventional soil sampling is far too costly to be routinely used on a large scale.

Specifically, \citet{ViscarraRossel2007} obtained 34,266 gamma-ray readings using a gamma-radiometer mounted on the front of a four-wheel-drive vehicle. After some preprocessing, they smoothed the data using ``local kriging'' and carried out a multivariate calibration of the hyperspectral gamma-ray data to predict soil properties. They showed that ``kriging improved the signal-to-noise ratio of the gamma-ray spectra.'' We focus here on spatial prediction of the total radioactivity count, the integrated count over the 0.4 - 2.81 mega-electronvolt spectrum, given in units of counts per second. The total count has been shown to be closely associated with the clay content in the soil \citep{Taylor2002,Pracilio2004}. Previously, \citet{Cressie2010} carried out an exploratory data analysis of total count and obtained spatial predictions using a spatial-random-effects model.

\begin{figure}
\centering\includegraphics[width=.95\textwidth]{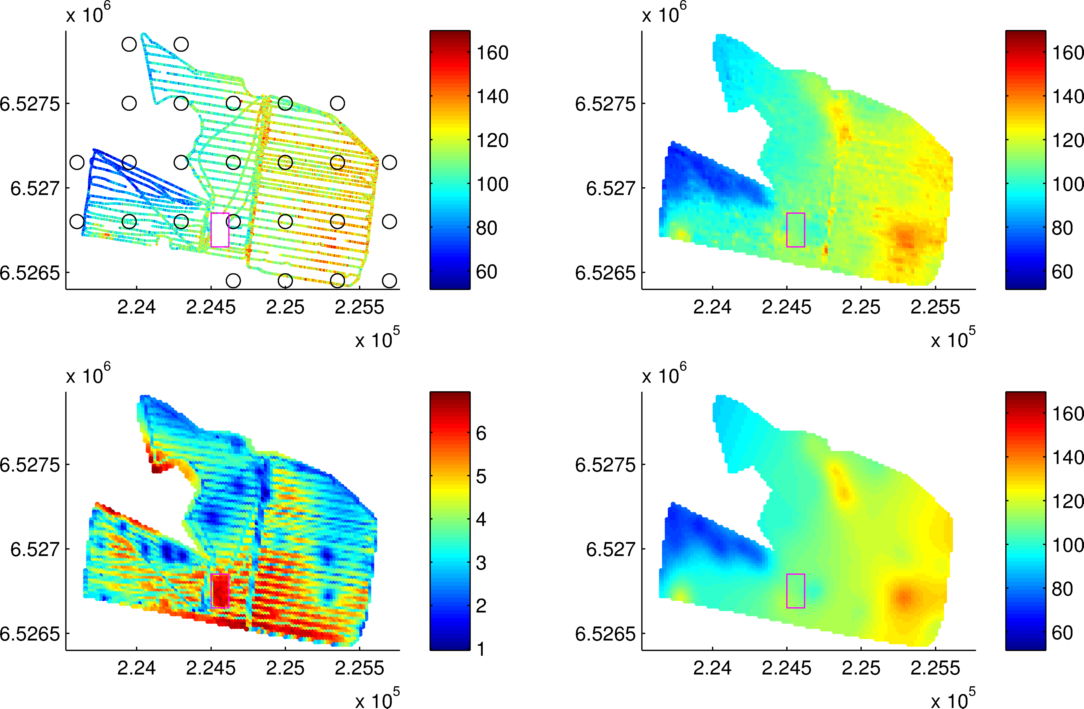}
\caption{\label{soilpred}Top left: Gamma emissions total count observations (small colored dots) and locations of the 25 basis-function centers for $\bb_\theta(\cdot)$ (black circles). Top right: Posterior mean of the true intensity using our model. Bottom left: Posterior standard deviation of the true intensity. Bottom right: Posterior mean of the smooth process without $\delta(\cdot)$ (see text). The test region (MBD) is represented by a pink rectangle. Color-scale units are counts per second; Easting and Northing are given in meters.}
\end{figure}

To assess prediction performance, we created a test region (called MBD) containing 409 observations.
The test data were only used for model evaluation, and they were not available for model fitting. The remaining $n = 33,866$ measurements, together with the test region (MBD), are shown in the top left panel of Figure \ref{soilpred}. The spatial domain was taken to be $\domain \colonequals (223525, 225770) \times (6526400, 6527930)$ in Easting and Northing.

Following \citet{Cressie2010}, we log-transform the (shifted) data to obtain additive measurement error:
\begin{equation}
\label{soiltransform}
Z(\bs_i) \colonequals \log(\textrm{TC}(\bs_i)+160), \; i=1,\ldots,n,
\end{equation}
where TC denotes the total radioactivity count.

\citet{Cressie2010} identified Easting and Northing as important trend terms, and so we set $\bx(\bs) \colonequals (1, \bs')'$, where each location $\bs \in \domain$ is a two-dimensional vector consisting of Easting and Northing (in meters). The measurement-error variance (on the log scale) is known from another experiment to have a value of $\sigma_\epsilon^2 = 0.0016$ \citep[see][]{Cressie2010}. As the empirical variance of $\bZ$ was calculated to be $\hat{\sigma}^2_Z= 0.0026$ (after subtracting the trend as estimated by ordinary least squares), the signal-to-noise ratio is less than 2. Nonetheless, it is possible to distinguish signal from noise in many areas of the domain due to high sampling density (see top left panel of Figure \ref{soilpred}).

We considered two equidistant grids of fixed knots on the domain $\domain$, one with 64 and one with 144 locations. For the vector $\bb_\theta(\cdot)$ in \eqref{generalcovparam}, we chose 25 power exponential functions with scale parameter 300 and centers shown in the top left panel of Figure \ref{soilpred}. We chose a tapering length of $L=35$ in \eqref{deltacovdef}, which resulted in roughly 150 nonzero elements per row for $\bV$ in \eqref{datacov} (i.e., about $99.56\%$ of the elements of $\bV$ were zero). The prior distributions of the parent-covariance parameters were as described in Section \ref{covariancesec}, with  $\mu_{\sigma}=\log ( \hat{\sigma}_Y )$ and $\mu_{\gamma}=\log ( 577.76 )$, where $\hat{\sigma}_Y \colonequals \sqrt{\hat{\sigma}^2_Z - \sigma^2_\epsilon}$.

On an Intel Xeon X5560 machine with 94.5 GB RAM, we ran an MCMC for each of the models for 20,000 iterations, of which 10,000 were considered burn-in (based on examination of trace plots), and we only used every 10th of the remaining iterations for inference. We also obtained the posterior distribution of $Y(\cdot)$ at a grid of 5,707 locations. In Figure \ref{soilpred}, using our model, we show the posterior means (top right panel) and standard deviations (bottom left panel) of the (error-free) true intensity (TI) on the original scale, defined in analogy to the transformation \eqref{soiltransform} as $\textrm{TI}(\bs) \colonequals \exp\{Y(\bs)\} -160$. We also show the posterior mean of $\exp\{\mu(\cdot) + \nu(\cdot) \} -160$ (i.e., without $\delta(\cdot)$) in the bottom right panel of Figure \ref{soilpred}.

The model comparison was carried out on the log-scale. We obtained samples from the posterior distribution of $Z(\bs_j)$ at test location $\bs_j$ as, $ Z^{(k)}(\bs_j) \colonequals Y^{(k)}(\bs_j) + \epsilon^{(k)}(\bs_j)$, where the $Y^{(k)}(\bs_j)$ are posterior samples from $Y(\bs_j)$, and $\epsilon^{(k)}(\bs_j) \sim N(0, \sigma^2_\epsilon)$ is independent ``measurement error.'' We then calculated the average squared distance (ASD) of the means of $ \{Z^{(k)}(\bs_j) \}$ to the test observations $Z(\bs_j)$, and the interval score for $95\%$ credible intervals for $Z(\bs_j)$, for all models, averaged over the test locations.

\begin{table}[ht]
\begin{center}
\caption{\label{soilresults_table}Summary of the results of the soil data analysis}
\begin{tabular}{| l | r r | r r | r r | }
\hline
& \multicolumn{2}{|c|}{Random knots} & \multicolumn{2}{|c|}{64 Fixed knots} & \multicolumn{2}{|c|}{144 Fixed knots}  \\
Parent covariance	&	\multicolumn{1}{c}{NPC}	&	\multicolumn{1}{c|}{SPC}	&	\multicolumn{1}{c}{NPC}	&	\multicolumn{1}{c|}{SPC}	&	\multicolumn{1}{c}{NPC}	&	\multicolumn{1}{c|}{SPC}	\\
\hline								
Time (hours)	&	89.87	&	95.05	&	59.02	&	59.15	&	158.84	&	152.36	\\
\hline
ASD (MBD) $\times 100$	&	0.26	&	0.28	&	0.27	&	0.28	&	0.32	&	0.30	\\
IS (MBD) $\times 100$	&	26.96	&	28.13	&	27.39	&	28.48	&	29.41	&	31.38	\\
\hline
Posterior mean of $r$ &  35.57 & 42.88 & (64) & (64) & (144) & (144)\\
\hline
\end{tabular}
\vspace{1pt}

NPC = nonstationary parent covariance; SPC = stationary parent covariance; ASD = average squared distance; IS = interval score; Time = total time for the MCMC; MBD = missing-by-design (test region)
\end{center}
\end{table}

The results are shown in Table \ref{soilresults_table}. Random knots resulted in lower average squared distance and interval score than fixed knots. With the exception of average squared distance for the models with 144 fixed knots, NPC also improved over SPC. More knots resulted in less accurate predictive distributions.

Based on examination of trace plots, mixing was somewhat slower for random knots than for fixed knots and slower for NPC than for SPC (both for the soil data here and for the simulated data in Section \ref{simstudy}). Because the same number of MCMC iterations was used for all models, the computation times given in Tables \ref{sim_results_sinefun}--\ref{soilresults_table} can be slightly misleading. However, since the focus in this article is not parameter estimation but on prediction, and predictive performance of the models was also assessed based on an equal number of MCMC iterations, we feel that the comparison is fair.

%%%%%%%%%%%%%%%%%%%%%%%%%%%%%
\section{Conclusions \label{conclusions}}

In this article, our starting point was the \citet{Sang2012} approach to analyzing large spatial datasets, which combines a low-rank predictive-process component with a tapered remainder component. To achieve enough flexibility for the nonstationary processes often encountered in real-world applications, we extended this model in two ways: First, the components in the model are parameterized based on a nonstationary Mat\'{e}rn parent covariance function, in which the parameters vary spatially according to linear combinations of spatial basis functions. Second, for the low-rank component, which can be written as a linear combination of spatial basis functions, we make inference on the number, locations, and shapes of the basis functions. Posterior inference via reversible jump MCMC and related issues are described in detail.

The results of a simulation study (Section \ref{simstudy}) and an analysis of a very large soil dataset (Section \ref{application}) indicate that the two extensions described above can result in improved predictive distributions, especially in terms of quantifying prediction uncertainty. We show that for (typically nonstationary) real-world processes, it should often \emph{not} be the goal to approximate a simple covariance model (e.g., the stationary Mat\'{e}rn covariance) as closely as possible. Results indicate that our model is sufficiently flexible to overcome a misspecified parent covariance, and its flexibility does not seem to result in a penalty in the unlikely event that the truth is, in fact, a simple stationary covariance (see Simulation Study 3 in Section \ref{simstudy}). Due to its adaptability, our model can be used to model highly nonstationary processes with varying levels of smoothness.

%%%%%%%%%%%%%%%%%%%%%%%%%%%%%%%%%%%
\section*{Acknowledgments}

This research was supported by NASA under grant NNH08ZDA001N issued through the Advanced Information Systems Technology ROSES 2008
Solicitation, and by the Mathematics Center Heidelberg. I would like to thank Huiyan Sang, Emily Kang, the editor, associate editor, two anonymous referees, and especially Noel Cressie for helpful advice and comments.
I am also grateful to James Taylor and Alex McBratney of the University of Sydney for making the Nowley soil dataset available. The collection of the data was directed by Professor McBratney and funded by the University of Sydney.

%%%%%%%%%%%%%%%%%%%%%%%%%%%%%

% \appendix{Further Simulation Experiments}

\bibliographystyle{apalike}

\end{document}